\documentclass[twocolumn,showpacs,amssymb,prd]{revtex4}

\usepackage{graphicx}
\usepackage{dcolumn}
\usepackage{bm}
\usepackage{epsfig}

\begin{document}
\newcommand{\gsim}{\hbox{\rlap{$^>$}$_\sim$}}
\newcommand{\lsim}{\hbox{\rlap{$^<$}$_\sim$}}


\title{Critical Test Of Gamma-ray Burst Theories}

\author{Shlomo Dado} \affiliation{Department of Physics and Space Research
Institute, Technion, Haifa 32000, Israel}

\author{Arnon Dar}
\affiliation{Department of Physics and Space Research
Institute, Technion, Haifa 32000, Israel}

\begin{abstract} 

Very long and precise follow-up measurements of the X-ray afterglow of 
very intense gamma-ray bursts (GRBs) allow a critical test of GRB 
theories. Here we show that the single power-law decay with time of the 
X-ray afterglow of GRB 130427A, the record-long and most accurately 
measured X-ray afterglow of an intense GRB by the Swift, Chandra and 
XMM Newton space observatories, and of all other known intense GRBs, is that 
predicted by the cannonball (CB) model of GRBs from their measured 
spectral index, while it disagrees with that predicted by the widely 
accepted fireball (FB) models of GRBs.
  
\end{abstract}


\maketitle 

\section{Introduction}
Gamma-ray bursts are brief flashes of gamma 
rays lasting between few milliseconds and several hours [1] from extremely 
energetic cosmic explosions [2].  They are usually followed by a 
longer-lived "afterglow" emitted mainly at longer wavelengths [3] (X-ray, 
ultraviolet, optical, infrared, microwave and radio). Roughly they fall 
into two classes, long duration ones (GRBs) that last more than $\sim$ 2 
seconds, and short hard bursts (SHBs) that typically last less than $\sim$ 
2 seconds [4]. The GRBs seem to be the beamed radiation emitted by highly 
relativistic jets [5] ejected in broad line supernova explosions of type 
Ic [6], following collapse of rapidly rotating stripped envelope high mass 
stars to a neutron star, quark star, or black hole. The origin of 
SHBs is not known yet, but it is widely believed to be highly relativistic 
jets presumably emitted in a phase transition in/of a compact star (white 
dwarf, neutron star or quark star) to a more compact state following 
cooling and loss of angular momentum, or in merger of compact stars in 
close binaries due to gravitational wave emission [7].

In the past two decades, two theoretical models of GRBs and their 
afterglows, the fireball (FB) model [8] and the cannonball (CB)  model 
[9], have been used extensively to interpret the mounting observational 
data on GRBs and their afterglows. Both models were claimed to describe 
successfully the observational data. But, despite their similar names, the 
two models were and still are quite different in their basic assumptions 
and predictions (compare, e.g., [10] and [11]). Hence, at most, only one 
of them can provide a correct physical theory of GRBs and their 
afterglows.

In the CB model [9], bipolar jets that are made of a succession of highly 
relativistic plasmoids (cannonballs) are assumed to be launched in 
accretion episodes of fall-back matter onto the newly formed compact 
object in broad-line SNeIc akin to SN1998bw. The gamma-ray pulses in a 
GRB are produced by inverse Compton scattering of glory light -the 
light halo formed around the progenitor star by scattered light from 
pre-supernova ejections- by the electrons enclosed in the CBs.  The 
afterglow is mainly synchrotron radiation emitted from the electrons of 
the external medium which are swept into the CBs and are accelerated there 
to very high energies by turbulent magnetic fields.

The FB models of GRBs evolved a long way from the original spherical FB 
models [8] to the current conical models [11] which assume that GRBs are 
produced by bipolar jets of highly relativistic thin conical shells 
ejected in broad line SNeIc explosions. In these models, the prompt 
emission pulses are synchrotron radiation emitted in the collisions 
between overtaking shells, while the continuous collision of the merged 
shells with the circumburst medium drives a forward shock into the medium 
and a reverse shock in the merged shells, which produce the synchrotron 
radiation afterglow.

The claimed success of both models to describe well the mounting 
observational data on GRBs and their afterglows, despite their complexity 
and diversity, may reflect the fact that their predictions depend on 
several choices and a variety of free parameters, which, for each GRB, are 
adjusted to fit the observational data. As a result, when successful 
fits to observational data were obtained, it was not clear whether they 
were due to the validity of the theory or due to the multiple choices and 
free adjustable parameters. Scientific theories, however, must be 
falsifiable [12]. Hence, only confrontations between accurate 
observational data and the key predictions of the GRB models, which do not 
depend on free adjustable parameters, can serve as critical tests of the 
validity of such models.

Critical tests of the origin of the prompt gamma-rays are provided, e.g., 
by their measured polarization, correlations between various prompt 
emission properties, and the GRB prompt emission energy relative to that 
of its afterglow. While the observations have confirmed the  
predictions of the CB model they have challenged those of the 
standard FB models [13].

Critical tests of the GRB theories are also provided by the observed GRB 
afterglow. In the FB model the origin of the afterglow is a forward shock 
in the circumburst medium driven by the ultra-relativistic jet, while in 
the CB model the afterglow is produced by the Fermi accelerated electrons 
which are swept into the jet. That, together with the different jet 
geometries, result in different falsifiable predictions for the afterglow 
light-curves. In particular, conical FB models predict a broken power-law 
decline of the light curve of the afterglow [14] where the pre-break 
temporal decline index $\alpha$ increases by $\Delta =3/4$ for an ISM 
like density distribution, or by $\Delta=1/2$ for a wind-like density 
distribution, independent of of the afterglow frequency [11]. The observed 
breaks in GRB afterglows, however, often are chromatic breaks with a 
break-time and $\Delta$ that depend on frequency and satisfy neither 
$1/2\leq\Delta \leq 3/4$ (see, e.g., FIG. 1) nor the FB closure relations 
[11]. E.g., an analysis of the Swift X-ray data on the 179 GRBs 
detected between January 2005 and and January 2007 and the optical AGs of 
57 pre- and post-Swift GRBs did not find any burst satisfying all the 
criteria of a jet break [15]. Moreover, many GRBs have afterglows that do 
not show any break at all. Consequently, it has been suggested that, 
perhaps, these 'missing jet breaks' take place at rather late-time, when 
the observations are not precise enough anymore or after they end [16].

Recently, however, the X-ray afterglow of GRB 130427A, the brightest 
gamma-ray burst detected by Swift [17] in the last 30 years, was followed 
with high precision by the sensitive X-ray observatories Chandra and XMM 
Newton for a record-breaking baseline longer than 80 million seconds [18], 
which allows a critical test of both the standard FB models and the CB 
model. Detailed comparison between the observed late-time X-ray 
afterglow of GRB 130427A, and that predicted by the standard fireball 
models has already been carried in [18]. It was concluded there that the 
forward shock mechanism of the standard FB models with plausible values 
for the physical parameters involved cannot explain the data, in both cases 
of constant density and stellar-wind circumburst media.

In contrast, in this paper we show that the observed X-ray afterglow of 
the very intense GRB 130427A that decays with time like a single power-law 
with no visible jet break until the end of the measurements, is that 
expected from the CB model for very intense GRBs, and its temporal decay 
index is precisely that expected in the CB model from its measured 
spectral index. Moreover, we show that, within errors, this is also the 
case for the late-time X-ray afterglows of all the 28 most intense GRBs 
with known redshift $z$, whose late-time afterglow was well measured.

\section {The X-ray Afterglow In The CB Model} 
The circumburst medium in front of a CB moving with a highly
relativistic  bulk motion Lorentz factor $\gamma\gg 1$ is completely 
ionized by the CB's radiation. In the CB's rest frame, the ions of the 
medium that are swept in generate within the CB  turbulent magnetic
fields whose energy density is assumed to be in approximate equipartition
with that of the impinging particles. The electrons that enter the CB with
a Lorentz factor $\gamma(t)$ in the CB's rest frame
are Fermi accelerated there and cool by emission of synchrotron radiation
(SR), which is isotropic in the CB's rest frame and has a smoothly broken
power-law. In the observer frame, the emitted photons are beamed into a
narrow cone along the CB's direction of motion by its highly relativistic
motion, their arrival times are aberrated, and their energies are boosted
by its Doppler factor $\delta$ and redshifted by the cosmic expansion 
during their travel time to the observer. For $\gamma^2\gg 1$ and a 
viewing angle $\theta^2\ll 1$ relative to the CB direction of motion, the 
Doppler factor satisfies $ \delta\approx 2\,\gamma/ [1+\theta^2\,\gamma^2]$.

The observed spectral energy density (SED) of the {\it unabsorbed}
synchrotron X-rays has the form (see, e.g., Eq.~(28) in [10])
\begin{equation}
F_{\nu} \propto n^{(\beta_x+1)/2}\,[\gamma(t)]^{3\,\beta_x-1}\,
[\delta(t)]^{\beta_x+3}\, \nu^{-\beta_x}\, ,
\label{Fnu}
\end{equation}
where $n$ is the baryon density of the external medium encountered 
by the CB at a time $t$ and  $\beta_x$ is
the spectral index of the emitted X-rays, $E\,dn_x/dE\propto E^{-\beta_x}$.

The swept-in ionized material decelerates the CB motion.
Energy-momentum conservation for such a plastic collision between 
a CB of a baryon number $N_{_B}$, a radius $R$ and an initial
Lorentz factor $\gamma(0)\gg 1$,
which  propagates in a constant density ISM at a redshift $z$,   
yields the deceleration law (Eq.~(4) in [19])
\begin{equation}
\gamma(t) = {\gamma_0\over [\sqrt{(1+\theta^2\,\gamma_0^2)^2 +t/t_d}
          - \theta^2\,\gamma_0^2]^{1/2}}\,,
\label{goft}
\end{equation}
where $t$ is the time in the observer frame since the
beginning of the afterglow,  and $t_d={(1\!+\!z)\, N_{_B}/ 8\,c\,   
n\,\pi\, R^2\,\gamma_0^3}$ is the deceleration time-scale.

For a constant-density ISM, Eqs.~(1) and (2) yield 
an afterglow  whose shape  depends only on  three parameters: 
the product $\gamma_0\,\theta$, the deceleration time scale $t_d$, 
and the spectral 
index $\beta_x(t)$. As long as $t\lsim t_b=(1+\theta^2\,\gamma_0^2)^2\,t_d$, 
$\gamma(t)$ and consequently also $\delta(t)$  change rather 
slowly with $t$ and generate a {\it plateau phase} of $F_\nu(t)$, 
which lasts  until $t\approx t_b$.
Well beyond $t_b$, Eq.~(2) yields 
$\delta(t)\approx\gamma(t)\propto t^{-1/4}$ and 
\begin{equation}
F_\nu(t)\propto [\gamma(t)]^{(4\,\beta_x+2)}\,\nu^{-\beta_x}
\propto t^{-\alpha_x}\, \nu^{-\beta_x}
\label{Fnulate}
\end{equation}
where  
\begin{equation}
\alpha_x=\beta_x+1/2.
\label{Relation}
\end{equation}
Such a canonical behavior of the X-ray afterglow of GRBs (which was 
predicted by the CB model [20] long before its empirical discovery with 
Swift [21]), is demonstrated in Figure 1, where the 0.3-10 keV X-ray 
light-curve of GRB 060729 that was measured with the Swift XRT [17] is 
plotted together with its best-fit CB model light-curve [10]. Its 
late-time afterglow between $1.5 \times 10^{5}- 
1.5\times 10^7$ s shows a power-law decline with $\alpha_x=1.46\pm 0.025$ 
[17]. In the the CB model, Eq.~(4) and the measured photon  index 
$\beta_x=0.99\pm  0.07$ [17]  yields   $\alpha_x=1.49\pm 0.07$, in good 
agreement with its observed value. Figure 1 also demonstrates that: (1) 
the observed fast decline $F_\nu \propto t^{-6}$
of the prompt X-ray emission is much steeper than that expected
in the FB model from high latitude emission with $F_x\sim 
t^{-(2+\beta_x)}\approx t^{-3}$ [11] for an observed $\beta_x\sim 1$, 
(2) the observed $\alpha_x\sim 0$ during the plateau phase does not 
satisfy the FB model pre-break closure relations, (3) the  
increases of $\alpha$ by $\Delta\approx 1.5$ beyond the break does not 
satisfy $0.5\leq \Delta\leq 0.75$, and (4) the closure relation 
of the standard FB model [11]  $\alpha=2\,\beta$ beyond the break is not 
satisfied.  
\begin{figure}[]
\centering
\epsfig{file=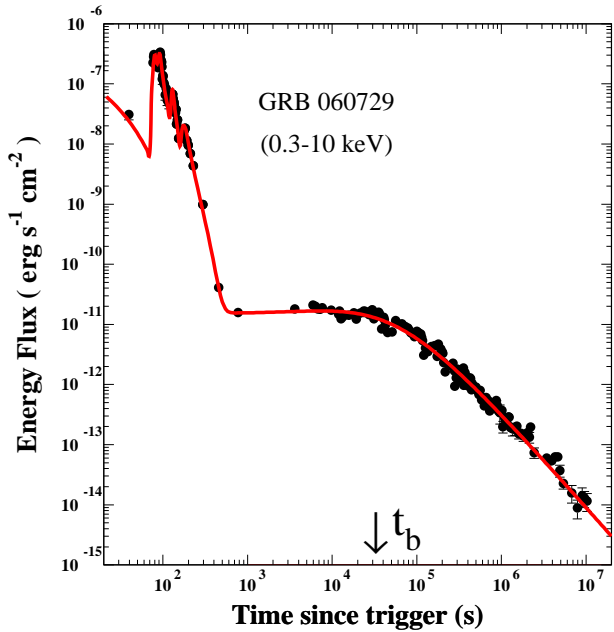,width=9.0cm,height=9.0cm}
\caption{The 0.3-10keV X-ray light-curve of GRB 060729 that was measured 
with the Swift XRT [17] and its best-fit CB model light-curve 
(Figure 1a from Ref. [10], updated). 
The late time afterglow as measured with the Swift XRT
between $1.5\times 10^5-1.5\times 10^7$ s had a spectral index  
$\beta_x=0.93\pm 0.07$ [17], which according to Eq.~(4)
yields $\alpha_x=1.43\pm 0.07$, in good agreement with  
the best fit late-time temporal decline index 
$\alpha_x=1.46\pm 0.025$.}
\label{fig1}
\end{figure}

\section{X-ray Afterglows With Missing Breaks} 
In the CB model, the break/bend time of the afterglow in the GRB rest 
frame satisfies [22] $t_b/(1+z)\propto 1/[(1+z)\,Ep\, Eiso]^{1/2}$, where 
$Eiso$ and $Ep$ are, respectively, the GRB equivalent isotropic gamma-ray 
energy and the observed peak photon energy.  Hence, very intense GRBs with 
relatively large $Ep$ and $Eiso$ values have a relatively small $t_b$, 
which can be hidden under the prompt X-ray emission or its fast decline 
phase [10]. Consequently, only the post break temporal decline of the 
afterglow with a decay index $\alpha_x=\beta_x+1/2$ is observed [10]. This 
is demonstrated in Figure 2 where the light-curve of the 0.3-10 keV X-ray 
afterglow of the very intense ($Eiso\approx 10^{54}$ erg [23]) GRB 061007
that was measured with the Swift XRT [17] is plotted together with the 
best-fit single power-law temporal decay index $\alpha_x=1.50 \pm 0.05$. 
This temporal index is in good agreement with $\alpha_x=1.51\pm 0.05$ 
predicted by Eq.~(4).
\begin{figure}[]
\centering
\epsfig{file=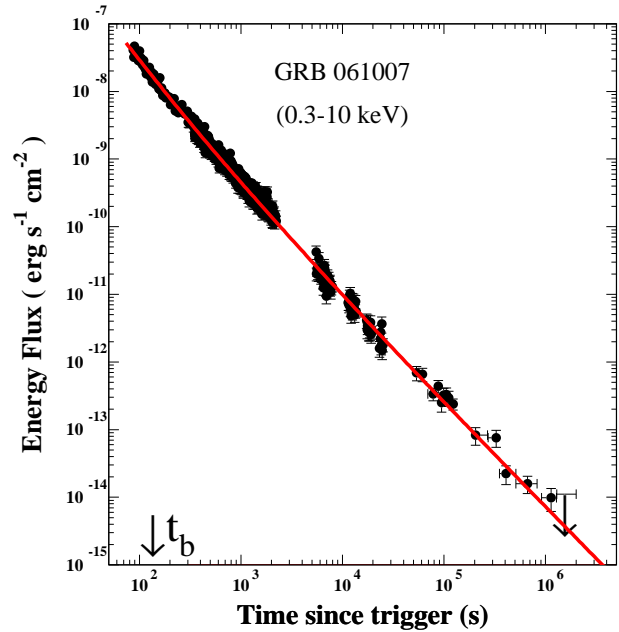,width=9.0cm,height=9.0cm}
\caption{The X-ray light-curve of the intense GRB 061007 that was 
measured with the Swift XRT [17] and 
its CB model best-fit (Figure 2a of Ref. [10], updated), which yields an 
early break and a start time 
hidden under the prompt emission phase, and a post break late-time 
temporal decay  index $\alpha_x=1.50\pm 0.05 $. The temporal decay 
index expected in the CB model from its measured spectral 
index $\beta_x=1.01\pm 0.05$ [24], as given by Eq.~(4), is 
$\alpha_x=1.51\pm 0.05$.} 
\label{fig2}
\end{figure}

To test further whether relation (4) is satisfied universally by the X-ray 
afterglow of the most energetic GRBs, we have extended our test to the 
X-ray afterglows of all GRBs with known redshift and $Eiso > 5\times 
10^{53}$ erg, which were followed up with an X-ray space based observatory 
for at least a few days, assuming a single power-law decline 
(corresponding to a constant ISM density along the CB trajectory). These 
GRBs are listed in Table 1 together with their measured redshift $z$, 
$Eiso$, temporal decay index $\alpha_x$ and spectral index $\beta_x$.  

The most energetic GRB listed in Table 1 is GRB 160625B, 
at redshift z=1.406, with $ Eiso\approx 5\times 10^{54}$ erg
measured by KONUS-Wind. In Figure 3, the light curve of 
its X-ray afterglow that was measured with the Swift XRT [17] is 
compared to its best fit single power-law light curve. The best fit
power-law has a
temporal decay index $\alpha=1.33\pm 0.04)$  in good agreement with the
expected value $1.33\pm 0.12$ 
from Eq.~(4) and the measured spectral index $\beta_x=0.83\pm 0.12$ [17].
\begin{figure}[]
 \centering
\epsfig{file=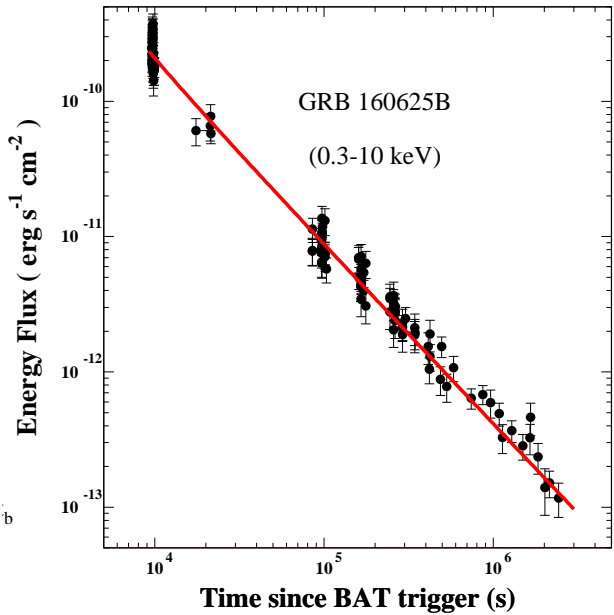,width=9.0cm,height=9.0cm}
\caption{The X-ray light-curve  of GRB 160625B, the most energetic 
GRB with known redshift ever detected by KONUS Wind,
as measured with the Swift XRT [17]. 
The line is the best-fit single power-law decay [17] with a temporal
index $\alpha_x=1.33\pm 0.04)$. The temporal decay index  
predicted by Eq.~(4) from  the measured  spectral index  
$\beta_x=0.83\pm 0.12$ [17] is $\alpha_x= 1.33\pm 0.12$.}
\label{fig3}
\end{figure}

In Figure 4, the measured values of $\alpha_x$ and  
$\beta_x$ for the 28 most intense GRBs 
with known redshift that are listed in Table 1
are compared to 
the CB model prediction (line) as given by Eq.~(4). 
The best-fit line $\alpha_x=a\,(\beta_x+1/2)$ 
to the data yields a=1.007. 
\begin{figure}[]
\centering
\epsfig{file=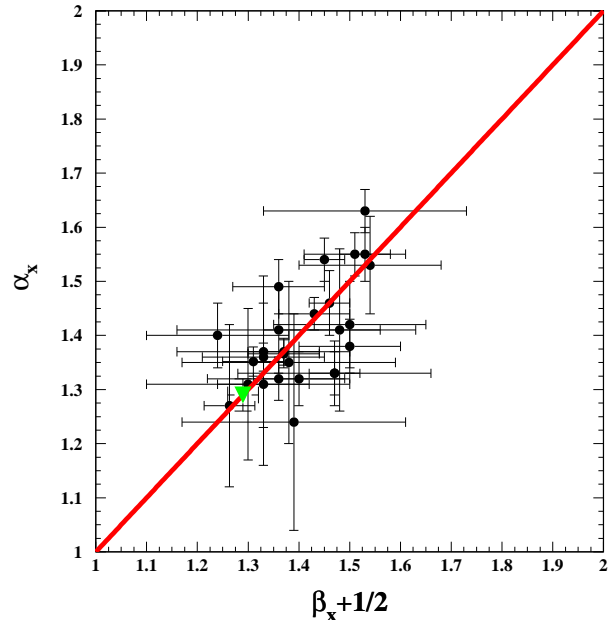,width=9.0cm,height=9.0cm} 
\caption{The values of the temporal index $\alpha_x$ as function of 
the spectral index $\beta_x+1/2$ for the 28 most intense GRBs 
with known redshift listed in Table 1 that were obtained from 
long follow-up measurements of their 0.3-10 keV X-ray afterglow with
Beppo-SAX, Swift XMM Newton and Chandra space based  observatories. 
The triangle indicates the value obtained for GRB 130427A.
The line is the CB model prediction, Eq.~(4).}
\label{fig4}
\end{figure}

\section{The X-ray Afterglow Of GRB 130427A}
The most accurate test, however, of the CB model relation  
$\alpha_x=\beta_x+1/2$ 
for a single GRB is provided by the follow-up measurements 
of the X-ray afterglow of GRB 130427A, the most intense GRB ever 
detected by Swift,  with the Swift XRT 
and with the sensitive X-ray observatories XMM Newton and Chandra up to a 
record time of 83 Ms after burst [17]. The measured light-curve  
has a single power-law decline with $\alpha_x = 1.309\pm 0.007$ in the 
time interval 47 ks - 83 Ms. 
The best single power-law fit to the combined measurements of the X-ray 
light-curve of GRB 130427A  with the Swift-XRT [17], 
XMM Newton and Chandra [18], and Maxi [25] that is 
shown in Figure 5 yields $\alpha_x=1.294\pm 0.03$.
The CB model prediction as given by  Eq.~(4,) 
with  the measured spectral index $\beta_x=0.79\pm 0.03$ [18],
is $\alpha_x=1.29\pm 0.03$ , in remarkable agreement with its best fit 
value.
\begin{figure}[]
\centering
\epsfig{file=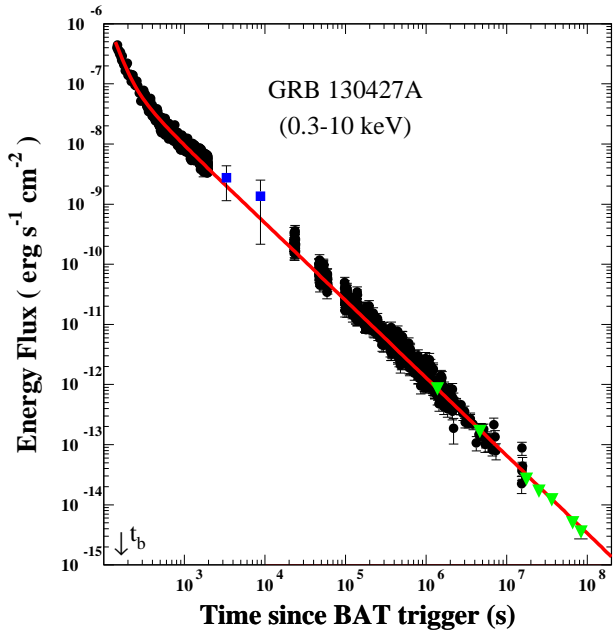,width=9.0cm,height=9.0cm}
\caption{The X-ray light-curve of the intense GRB 130427A that was
measured with Swift XRT [17] (circles)
and with  XMM Newton  and  Chandra [18] (triangles) up
to 83 Ms after burst, and its CB model best-fit with a start time and an
early break hidden under the prompt emission phase.
Also shown are the two MAXI data points [25] (squares)
at t = 3257 s and t = 8821 s. The best-fit 
power-law decline has an index $\alpha_x=1.29$. The temporal decay index 
predicted by the CB model, Eq.~(4), for the measured spectral index [18] 
$\beta_x=0.79\pm 0.03$ is $\alpha_x =1.29\pm 0.03$.} 
\label{fig5}
\end{figure}

No doubt, the assumptions of a constant density circumburst medium 
is an over simplification: Long duration GRBs are produced 
in supernova explosions of type Ic of short-lived massive stars, 
which take place mostly in superbubbles formed by star formation.
Such superbubble environments may have a bumpy density, 
which deviates significantly from the assumed  constant-density ISM.
Probably it is responsible for the observed deviations from the predicted 
smooth light-curves and $\chi^2/df$ values slightly larger than 1.  
Moreover, in a constant-density ISM, 
the late-time distance of a CB from its launch point 
is given roughly by,  
\begin{equation}
x={2 c \int^t \gamma\delta dt\over 1+z}\approx 
{8\,c\,\gamma_0^2\sqrt{t_d\,t}\over {1+z}}\,.    
\end{equation}
It may exceed the size of the superbubble 
and even the scale-height of the disk of the GRB 
host galaxy. In such cases, the transition of a CB from the 
superbubble into the Galactic ISM, or into the Galactic halo 
in face-on disk galaxies, will bend the 
late-time single power-law decline 
into a more rapid decline, depending on the density profile above the 
disk. Such a behavior  may have been observed by the Swift 
XRT [17] in a few  GRBs, such as 080319B and 110918A, at 
$t> 3\times 10^6$ s and in GRB 060729 at $t>3\times 10^7$ s by Chandra [26].

\section{Discussion and conclusions} The 83-Ms long follow-up measurements 
with the Swift XRT and the sensitive Chandra and XMM Newton observatories 
[18] of the X-ray afterglow of GRB 130427A, the brightest gamma-ray burst 
detected in the last 30 years, , allowed the most accurate test so far of 
the main falsifiable predictions of the standard FB models for the X-ray 
afterglow of GRBs. These predictions are a broken power-law light-curve 
with a late-time achromatic break, a post break temporal decay index 
larger by $1/2\leq \Delta\leq 3/4$ than its pre-break value, and closure 
relations between the temporal decay index and the spectral index of the 
afterglow for both pre-break and post break times. The precise record-long 
measurements of the X-ray afterglow of GRB 130427A disagree with these 
predictions of the standard FB models where a conical jet drives a forward 
shock into the circumburst medium [18]. In particular, the closure 
relations predicted by the fireball model require far-fetched values for 
the physical parameters involved, in both cases of constant density and a 
wind-like circumburst medium [18].

In contrast, the observed temporal decline like a single unbroken 
power-law of the light-curve of the 0.3-10 keV X-ray afterglow of 
GRB 130427A is that predicted by the CB model for the measured 
spectral index of its afterglow. In the CB model, the X-ray afterglow has 
a deceleration break that takes place at a time $t_b$ after the beginning 
of the afterglow (not necessarily the beginning of the GRB), and satisfies 
the correlation $t_b/(1+z)\propto 1/ [(1+z)\, Ep\, Eiso]^{1/2}$ [23]. 
Consequently, 
in very intense GRBs, the break is often hidden under the prompt emission 
or its fast decline phase. For GRB 130425A at z=0.34, with Eiso$\approx 
8.5\times 10^{53}$ erg and $Ep\approx 1200$ keV [27], the above 
correlation [20] yields a deceleration break at $t<200$ s, which, 
probably, was hidden under the fast declining phase of the prompt emission 
(see Figure 5).

Moreover, most of the X-ray afterglows of the 28 most intense GRBs among 
the GRBs with known redshift that were followed long enough with one or 
more of the space based X-ray telescopes Beppo-SAX, Swift, Chandra and 
XMM-Newton, have light-curves $F_\nu(t) \propto t^{-\alpha_x}\, 
\nu^{-\beta_x}$ with temporal and spectral indices that satisfy within 
errors the relation $\alpha_x=\beta_x+1/2$ predicted by the CB model for a 
constant density circumburst medium.

Furthermore, in the FB models, the predicted achromatic break in the light 
curve of the X-ray afterglow of GRBs is a direct consequence of the 
assumed conical geometry of the the highly relativistic jet - a conical 
shell with a half opening angle $\theta_j\gg 1/\gamma_0$, where 
$\gamma_0\gg 1$ is the initial bulk motion Lorentz factor of the jet. The 
failure of the conical fireball models to predict correctly the observed 
break properties in GRB afterglows, and the absence of a jet break in the 
X-ray afterglow of very intense GRBs such as 130427A, probably, is due to 
the assumed conical geometry. This is supported by the fact that, unlike 
the cannonball model, the conical fireball models have failed to predict 
other major properties of GRBs which strongly depend on the assumed 
conical jet geometry. That includes the failure to predict/reproduce the 
observed canonical shape of the lightcurve of the X-ray afterglow of GRBs 
[21] and the main properties of its various phases: The rapid spectral 
softening during the fast decline phase of the prompt emission, which was 
interpreted in the framework of the conical FB models as high latitude 
emission [28], was not expected/predicted. The plateau phase that follows 
was not reproduced and was interpreted aposteriory by postulating 
continuous energy injection into the blast wave by hypothetical central 
GRB engines, such as magnetars [29]. Furthermore, unlike the cannonball 
model, where X-ray flashes (XRFs) and low-luminosity GRBs were 
successfully explained as GRBs produced by SNeIc akin to SN1998bw and 
viewed far off axis [30], the collimated fireball models could not explain 
why GRBs such as 130427A and 980425, which were produced by the very 
similar broad line stripped envelope SN2013c and SN1998bw, respectively 
[31], have isotropic equivalent energies which differ by six orders of 
magnitude. Moreover, the GRB/SN association and the short lifetime of of 
the massive stars which produce SNeIc imply that the rates of GRBs and 
star formation are related. But, while the cannonball model predicted 
correctly the redshift distribution of the joint population of GRBs and 
XRFs from the observed dependence of the star formation rate on redshift 
(32), the conical fireball model did not (33).

\begin{table}
\caption{The temporal decay index $\alpha_x$ and  the 
spectral index  $\beta_x$ of the late-time  0.3-10 keV X-ray afterglow 
of the 28 most intense GRBs ($Eiso>0.5\times 10^{54}$ erg) with known 
redshift and long follow-up afterglow measurements 
with Beppo-SAX, Swift, Chandra and XMM-Newton.}
\begin{tabular}{l l l l l l}
\hline
\hline
GRB &$~~$z & $~~$Eiso  & $~~~~\alpha_x$ & $~~~~\beta_x$ &  \\
    &   &$10^{54}erg$ &         &                & \\ 
\hline
990123 & 1.6    & 2.78 & $1.46 \pm.06$  &  $0.96\pm .04$  \\
010222 & 1.477  & 1.14 & $1.33 \pm .04$ &  $0.97 \pm .05$ \\
061007 & 1.26   & 1.0  & $1.55 \pm .05$  & $1.03 \pm .05$ \\
070328 &2.0627  & 0.64 & $1.44 \pm .03$ &  $0.93 \pm .07$ \\
080607 & 3.036  & 1.87 & $1.53 \pm .09$ &  $1.04 \pm .14$ \\
080721 & 2.591  & 1.21 & $1.49 \pm .05$ &  $0.86 \pm .09$ \\
080810 & 3.35   & 0.5  & $1.42 \pm .08$ &  $1.00 \pm .15$ \\
080916C& 4.35   & 0.88 & $1.31 \pm .14$ &  $0.80 \pm .20$ \\
090323 & 3.57   & 3.98 & $1.35 \pm .15$ &  $0.88 \pm .21$ \\
090423 & 8.26   & 0.89 & $1.41 \pm .08$ &  $0.86 \pm .20$ \\
090812 & 2.452  & 0.44 & $1.32 \pm .04$ &  $0.86 \pm .14$ \\
090902B& 1.822  & 3.6  & $1.40 \pm .06$ &  $0.74 \pm .14$ \\
090926A& 2.1062 & 2.0  & $1.41 \pm .05$ &  $0.98 \pm .10$ \\
110205A& 2.22   & 1.36 & $1.55 \pm .04$ &  $1.01 \pm .10$ \\
110422A& 1.77   & 0.72 & $1.32 \pm .05$ &  $0.90 \pm .09$ \\
110731A& 2.83   & 0.46 & $1.26 \pm .04$ &  $0.76 \pm .05$ \\
110918A& 0.984  & 2.11 & $1.63 \pm .04$ &  $1.03 \pm .19$ \\
130427A& 0.3399 & 0.85 & $1.29 \pm .03$ &  $0.79 \pm .03$ \\
130505A& 2.27   & 3.8  & $1.27 \pm .15$ &  $0.76 \pm .05$ \\
131108A& 2.4    & 0.58 & $1.33 \pm .06$ &  $0.97 \pm .19$ \\
140419A& 3.956  & 1.9  & $1.37 \pm .03$ &  $0.87 \pm .07$ \\
140206A& 2.73   & 2.4  & $1.29 \pm .03$ &  $0.80 \pm .06$ \\
150206A& 2.087  & 0.6  & $1.25 \pm .03$ &  $0.79 \pm .07$ \\
150314A& 1.758  & 0.69 & $1.53 \pm .04$ &  $0.95 \pm .04$ \\
150403A& 3.139  & 0.6  & $1.37 \pm .14$ &  $0.83 \pm .17$ \\
151021A& 2.330  & 1.0  & $1.38 \pm .05$ &  $1.00 \pm .10$ \\
160131A& 0.972  & 0.83 & $1.24 \pm .20$ &  $0.89 \pm .22$ \\
160625B& 1.406  & 5.0  & $1.34 \pm .05$ &  $0.83 \pm .12$ \\  
\hline
\end{tabular}
\end{table}

{\bf Acknowledgment}: We thank an anonymous referee for 
useful comments and suggestions.

\end{document}